\documentclass[11pt,twoside]{article}
\usepackage{asp2004}
\usepackage{psfig}
\usepackage{epsf}
\usepackage{graphics}
\usepackage{lscape}
\begin{document}
\title{Dynamics of Intermediate Mass Black Holes in Star Clusters}
\author{H. Baumgardt}
\affil{Institute of Advanced Physical and Chemical Research  RIKEN,
   2-1 Hirosawa, Wako-shi, Saitama 351-019, Japan}

\author{S. F. Portegies Zwart}
\affil{Astronomical Institute `Anton Pannekoek',
                University of Amsterdam, Kruislaan 403}
\affil{Institute for Computer Science,
                University of Amsterdam, Kruislaan 403}

\author{S. L. W. McMillan}
\affil{Department of Physics, Drexel University,
                Philadelphia, PA 19104, USA}

\author{J.  Makino}
\affil{Department of Astronomy, School of Science, The University of Tokyo,
            7-3-1 Hongo, Bunkyo-ku, Tokyo 113-0033, Japan}

\author{T. Ebisuzaki}
\affil{Institute of Advanced Physical and Chemical Research  RIKEN,
   2-1 Hirosawa, Wako-shi, Saitama 351-019, Japan}

\begin{abstract}
We have followed the evolution of multi-mass star clusters containing massive central black holes
by N-body simulations on the GRAPE6 computers of Tokyo University.
We find a strong cluster expansion and significant structural changes
of the clusters. Clusters with IMBHs have power-law density profiles
$\rho \sim r^{-\alpha}$ with slopes $\alpha=1.55$ inside the influence 
sphere of the central black hole. This leads to a constant
density profile of bright stars in projection, which rules out the presence 
of intermediate mass black holes in core collapse clusters. If the star clusters
are surrounded by a tidal field, a central IMBH speeds up the destruction of the
cluster until a remnant of a few hundred stars remains, which stays bound
to the IMBH for a long time.  We also discuss the efficiency of different
detection mechanisms for finding IMBHs in star clusters.
\end{abstract}
\thispagestyle{plain}

\section{Introduction}

X-ray observations of starburst and interacting galaxies have revealed a class of ultra-luminous X-ray
sources (ULX), with luminosities of order $L\approx 10^{39}$ to $10^{41}$ \citep{Makishimaetal2000}. 
If the flux is radiated isotropically, this exceeds the Eddington luminosities of stellar mass black holes
by orders of magnitude, making ULX good candidates for IMBHs.
Many ULX appear to be associated with star clusters \citep{Fabbianoetal1997},
the irregular galaxy M82 for example hosts an ULX with luminosity
$L > 10^{40}$ erg/sec near its center \citep{Matsumotoetal2001, Kaaretetal2001} whose position
coincides with that of the young ($T \approx 10$ Myrs) star
cluster MGG-11. \citet{PortegiesZwartetal2004a} and \citet{McMillanetal2004} have performed $N$-body 
simulations of several star clusters in M82 and showed that runaway merging 
of massive stars could have led to the formation of an IMBH with a few hundred to a few thousand
solar masses in MGG-11, thereby explaining the presence of the ultraluminous X-ray source.

The fact that a considerable fraction of star cluster might have formed intermediate mass black holes (IMBHs) has 
interesting consequences. For example, IMBHs of a few 100 to a few 1000 $M_\odot$ would explain 
why the mass-to-light ratios in several globular clusters increase towards the center \citep{Gerssenetal2002, 
Colpietal2003}, although the data presented so far is also compatible with an unseen concentration of 
neutron stars and heavy mass white dwarfs \citep{Baumgardtetal2003}. IMBHs in star clusters would also be
prime targets of the forthcoming generation of ground and space-based gravitational wave detectors
and could provide the missing link between the stellar
mass black holes formed as the end product of stellar evolution and the $10^6$ to $10^9 M_\odot$
sized black holes found in galactic centers \citet{Ebisuzakietal2001}.

In this paper we explore the dynamical evolution of star clusters containing massive black holes. We study how
star clusters evolve during a Hubble time and compare the outcome of our simulations with galactic globular
clusters in order to determine which clusters are likely to
contain IMBHs.  We also study what is left bound to an IMBH after the parent cluster is dissolved and 
discuss ways how to detect an IMBH in a globular cluster.

\section{Details of the Simulations}

We simulated the evolution of star clusters containing between $N = 16,384$ (16K) and $131,072$ (128K) stars,
using the collisional Aarseth $N$-body code NBODY4 \citep{Aarseth1999} on the GRAPE6 computers of Tokyo
University\linebreak \citep{Makinoetal2003}. Clusters were treated as isolated and followed King $W_0=7.0$
profiles initially. The models started with IMBHs of $M_{BH} = 1000 M_\odot$ that were
initially at rest in the cluster centers. When creating the clusters, stellar velocities were chosen  
such that the initial model was in dynamical equilibrium in the combined potential
of the cluster and the central IMBH.
Our simulations included stellar evolution, dynamical relaxation and the tidal disruption of stars 
which get too close to the central black hole. The initial half-mass radius of the clusters was 4.9 pc. So far
we have not included stellar collisions or the change of the stellar orbits due to gravitational radiation
into the code since these processes are not likely to play an important role for the type of clusters
considered in this study. 

The mass function of stars was given by a \citet{Kroupa2001}
IMF with lower mass limit of $0.1 M_\odot$ and we modelled stellar evolution by the fitting
formulae of \citet{Hurleyetal2000}. 
Two series of simulations were made, one 
with a mass-function extending up to $30 M_\odot$ and a second series in which the maximum stellar
mass was equal to $100 M_\odot$. In the first series, only few black holes were formed, all of them
with masses below $3 M_\odot$, while in the second series a significant number of black holes
with masses up to $45 M_\odot$ were formed. We assumed a 100\% retention rate for black holes in the 
clusters at time of formation, so the situation in real globular clusters is probably somewhere between
our two cases. More details of the simulations can be found in \citet{Baumgardtetal2004b}. 

\section{Results}

\subsection{Density profiles}

\begin{figure}[t!]
\begin{center}
\plotone{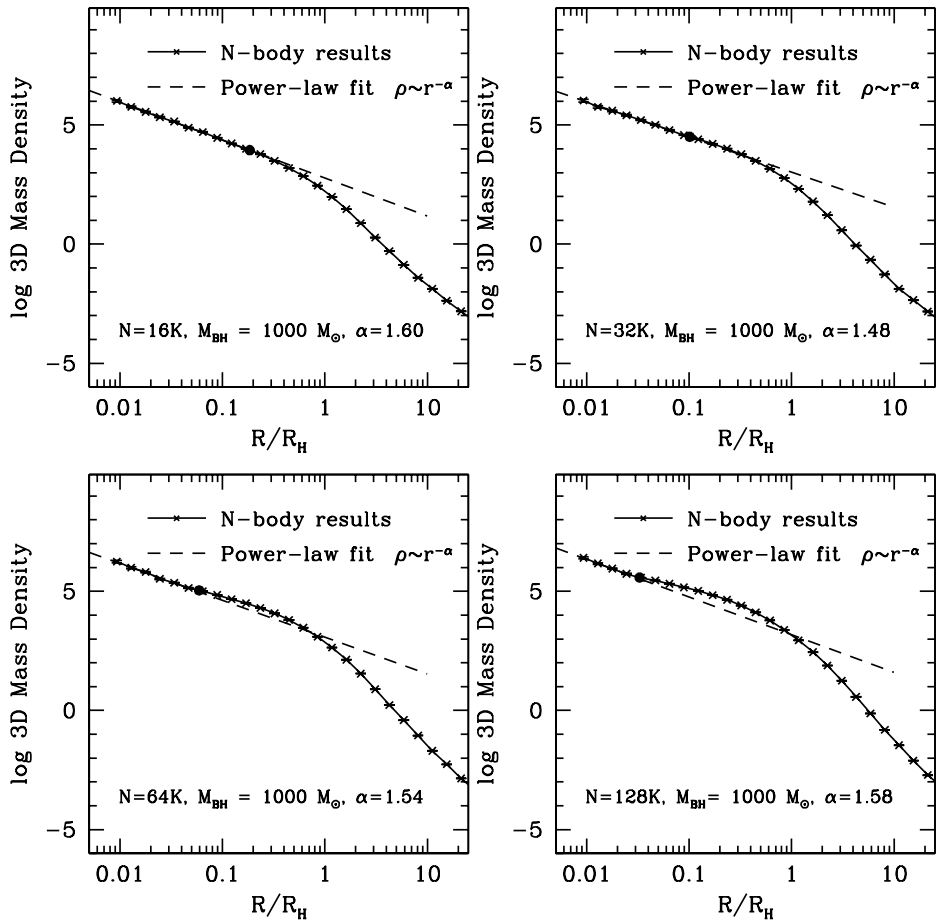}
\end{center}
\caption{3D mass density profile after T=12 Gyrs for 4 clusters starting with particle numbers between
 $16,384 \le N \le 131,072$. Solid lines mark the $N$-body results, dashed lines
  a single power-law fit to the density profile inside the influence radius of the black hole (shown by
   a solid circle). For all models we obtain slopes near $\alpha=1.55$ for the central stellar cusp.}
\label{3d_dens}
\end{figure}

Fig.\ \ref{3d_dens} depicts the final density profile of the four clusters with different particle
numbers after 12 Gyrs. Shown is
the three-dimensional mass density of all stars. In order to calculate the density profile, we overlayed
between 5 (128K) to 20 (16K) snapshots centered at T=12 Gyrs, creating roughly the same statistical
uncertainty for all models. All snapshots were centered on the position of the IMBH.
We then fitted the combined density profile inside the influence radius of the black hole
with a power-law density profile. As can be seen, we obtain power-law profiles $\rho \sim r^{-\alpha}$ 
inside the influence radius of the black hole with a slope around $\alpha=1.55$ for all clusters.
There is no dependence of the slope on the particle number.
\begin{figure}[t!]
\plotone{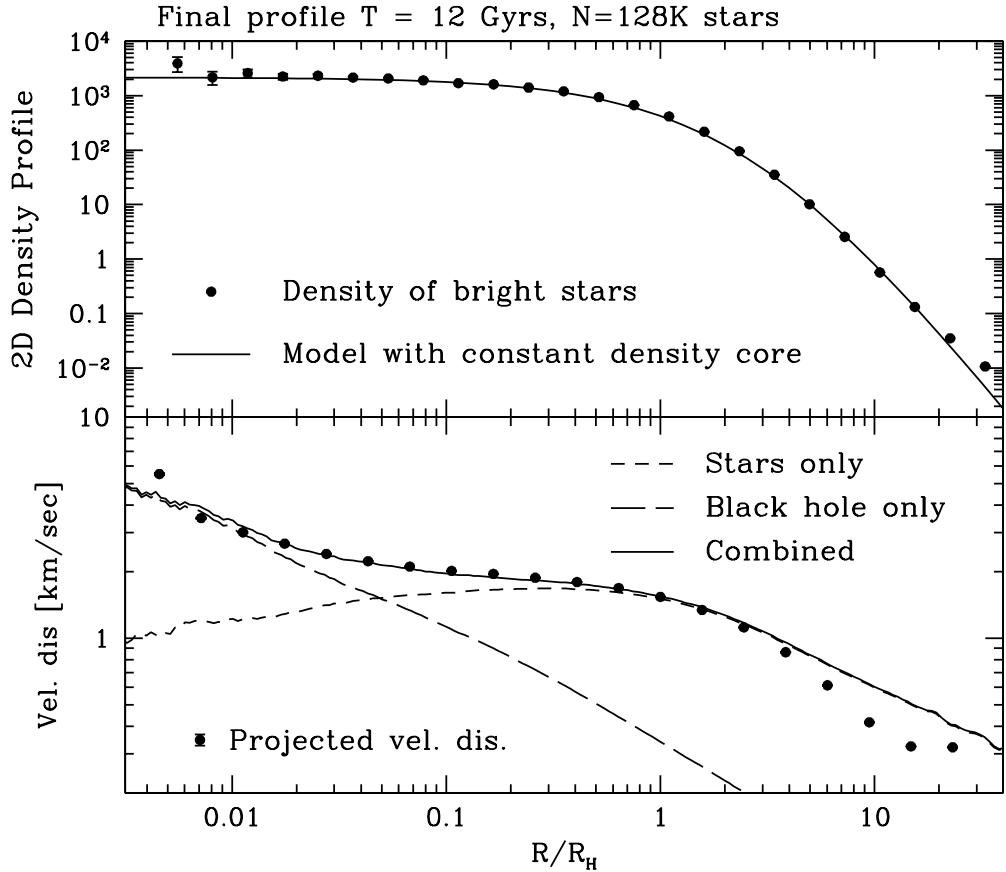}
\caption{Projected density profile of bright stars (top) and projected velocity dispersion of the cluster
starting with $N=131,072$ stars. The projected distribution of bright stars has a constant
density core, similar to that seen in most globular clusters.
Observations of the velocity dispersion could reveal the black hole if a sufficiently large number of
stars at radii $r/r_h<0.01$ can be observed (bottom panel).}
\label{2d_dens}
\end{figure}

The slope we obtain for multi-mass clusters is slightly flatter than the $\alpha=1.75$ slope found for 
single-mass clusters by \citet{BahcallWolf1976} and \citet{Baumgardtetal2004a}. 
The reason is that while high mass stars 
still follow an $\alpha=1.75$ profile, they are not numerous enough to determine the overall profile.
The upper panel of Fig. \ref{2d_dens} depicts the projected distribution of bright stars for the cluster
with $N=128$K
stars. We define bright stars to be all stars with masses larger than 90\% of the turn-off mass 
which are still main-sequence stars
or giants at $T=12$ Gyrs. Their density distribution should be representative of the distribution of 
cluster light.
The projected density distribution of bright stars does not show a central rise and can instead be fitted
by a model with a constant density core. The reason is that due to mass segregation, compact remnants,
which are more massive than main-sequence stars, have been enriched in the core while the density of
main-sequence stars has decreased
in the center. A cluster with a massive central black hole would therefore
appear as a standard King profile cluster to an observer, making it virtually indistinguishable from
a star cluster before core collapse.
Core collapse clusters have power-law density profiles in their centers, which is in contradiction with 
this profile. Since the central relaxation times of core collapse
clusters are much smaller than a Hubble time, any cusp profile would have been transformed into a
constant density core if an IMBH would be present in any of these clusters, so the presence of IMBHs
in core collapse clusters is ruled out.

The lower panel of Fig.\ \ref{2d_dens} shows the velocity dispersions, both
the measured one and the one inferred from the mass distribution of stars.
The inferred velocity dispersions were calculated from Jeans equation (Binney \& Tremaine 1986, eq.\ 4-54)
and different mass distributions under the assumption
that the velocity distribution is isotropic (i.e. $\beta=0$). The velocities calculated from the mass
distribution of the cluster stars alone give a good fit at
radii $r/r_h>0.2$ where the mass in stars is dominating
(except at the largest radii, where the velocity distribution becomes radially anisotropic).
At radii $r/r_h<0.2$, the contribution of the black hole becomes important. At a radius $r/r_h = 0.01$, the
velocity dispersion is already twice as high as the one due to the stars alone. For a
globular cluster at a distance of a few kpc, such a radius corresponds to central distances of one
or two arcseconds. Of order 20 stars
would have to be observed to detect the central rise at this radius with a 95\% confidence limit. This
seems possible both for radial velocity or proper motion studies with HST.

\subsection{Gravitational radiation}

\begin{figure}[t!]
\plotone{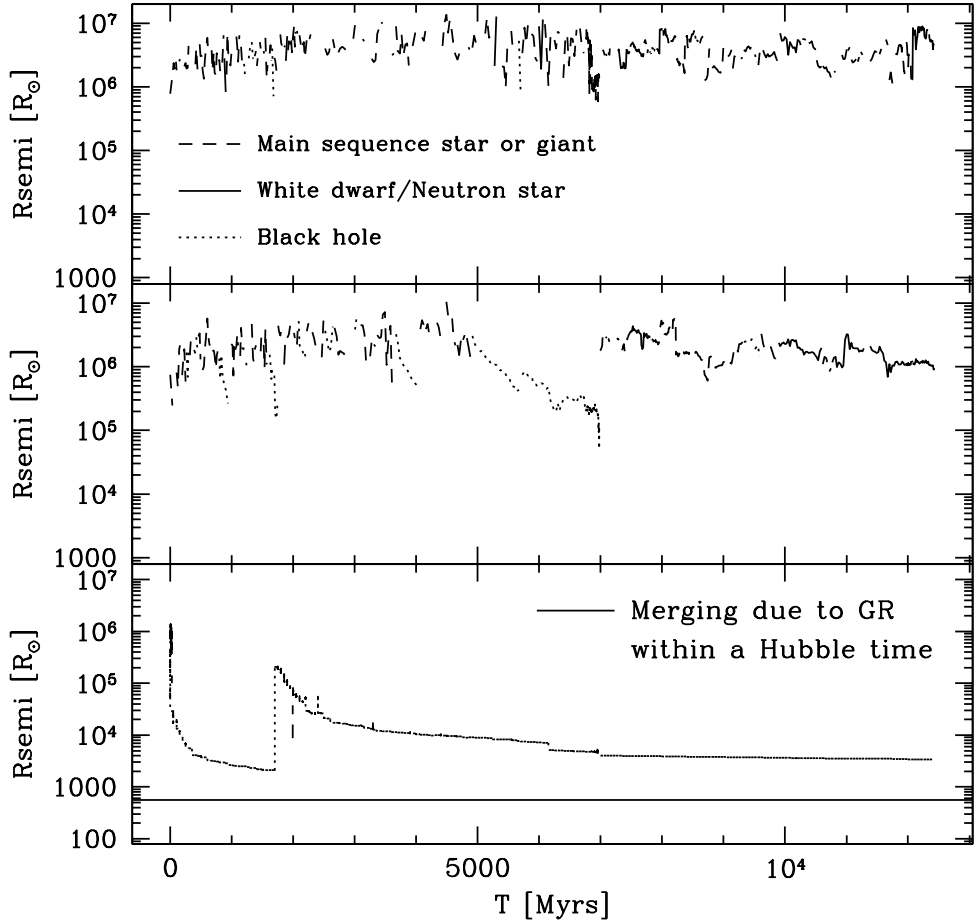}
\caption{Semi-major axis of the three stars deepest bound to the IMBH as function of time for the cluster
with $N=128$K stars and a high upper mass limit. The star closest bound to the IMBH is almost always another black hole
which is among the heaviest stars in the cluster. The other stars are too far away from the IMBH to undergo 
mass transfer.}
\label{near_u}
\end{figure}

Fig.\ \ref{near_u} depicts the semi-major axis of stars which are deepest bound to the IMBH
for a cluster with $N=128$K stars and a mass-function that extends up to 100 $M_\odot$. The energy of the 
deepest bound star decreases quickly in the beginning
when it still has many interactions with passing stars. When the semi-major axis becomes significantly smaller 
than that of other deeply bound stars, interactions become rare and the energy change slows down
considerably. For all simulations made, the innermost
stars are among the heaviest stars formed in the cluster and would be a massive black holes with several
$10 M_\odot$ for
an IMBH in a globular cluster. The innermost star will therefore not transfer mass onto the IMBH.
All other stars have
semi-major axis of $R>10^6 R_\odot$ which is too far for mass transfer, even if some stars will move on strongly
radial orbits. An IMBH in a star cluster can therefore only accrete gas from disrupted stars, or when   
a star captured through tidal heating is close enough to the IMBH to undergo mass transfer \citep{Hopmanetal2004}.

The dashed line in Fig.\ \ref{near_u} marks the radius inside which a $20 M_\odot$ black hole can merge with
a 1000 $M_\odot$ IMBH within a Hubble time. The orbit of the deepest bound star is still a factor of 6  
wider than this radius, so gravitational radiation does not significantly change the stellar orbit.
If the number of cluster stars is higher or the initial model more concentrated, the innermost  
star would be bound more tightly and the two black holes could merge with each other. 
In this case the system would become visible for gravitational wave telescopes like {\it LISA} during
the final stages before merging.

\subsection{Effects of a tidal field}

\begin{figure}[t!]
\plotone{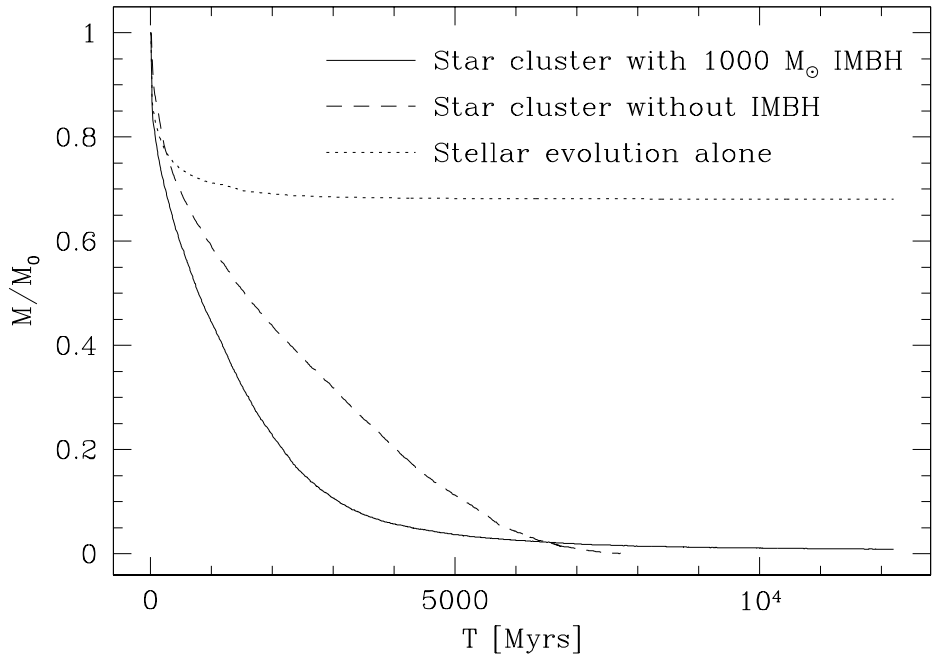}
\caption{Bound mass as a function of time for two clusters with $N=16$K stars moving in circular orbits
with $R_G=8$ kpc around the galactic center. The cluster with a 1000 $M_\odot$ IMBH in its center (solid line) 
dissolves about 
twice as fast as the one without an IMBH. A cluster remnant of about 100 stars remains bound to the central IMBH
for a long time.}
\label{tidf}
\end{figure}

Fig.\ \ref{tidf} shows the mass of bound stars as a function of time for two $N=16$K clusters, one with an IMBH 
of $1000 M_\odot$ in its center and one without an IMBH. Both clusters move in circular orbits with radius
$R=8$ kpc around the galactic center. The bound mass decreases in both clusters due to mass loss from stellar 
evolution and since during each relaxation time a certain fraction of stars gains the energy necessary for escape 
through encounters with
other cluster stars. The cluster with an IMBH loses its mass even faster since in addition to the previous
processes, tidal disruption of stars by the IMBH also decreases the average energy of
the cluster stars, thereby heating the whole system. As a result, stars flow over the tidal boundary much
faster.
Mass loss slows down considerably when the number of stars has dropped to less than a few hundred stars, since 
by then most mass is in the central black hole and the 
relaxation time starts to increase with decreasing cluster mass.
As a consequence, a system of about 100 stars, composed mainly out of main sequence stars and white dwarfs,
is still bound to the IMBH after a Hubble time. In the solar neighbourhood, 
the central IMBH could easily be found in such a cluster through kinematic studies,
since the mass-to-light ratio is very high and the stellar velocities show a near perfect Keplerian rise. 
Near 
the galactic center, such clusters would spiral into the galactic center through dynamical friction 
\citep{PortegiesZwartetal2004b}. 
If the cluster does not contain a central black hole, it is likely to
be disrupted before it reaches close enough to the center. However, the
presence of the IMBH prevents the complete disruption of the cluster.
The innermost stars would be stripped
from the IMBH only in the very late stages, which could explain the presence of a group of young, massive
main-sequence stars less than 0.1 pc from the galactic center black hole \citep{HansenMilosavljevic2003}.

\section{Conclusions}

We have performed two sets of $N$-body simulations of multi-mass star clusters
containing intermediate mass black holes. We found that the 3-dimensional mass-density follows
a $\rho \sim r^{-1.55}$ density profile around the central black hole.
When viewed in projection, the luminosity profile of clusters with massive black holes has a constant density
core. The presence of intermediate mass black holes in core collapse globular clusters like M15 is
therefore ruled out by our simulations. As was shown in \citet{Baumgardtetal2003}, a more natural explanation
for mass-to-light ratios that increase towards the center in such clusters is a dense concentration of
neutron stars, white dwarfs and stellar mass black holes.
The detection of a central black hole through proper motion or radial velocity measurements of stars
in the central cusp around the black hole is possible with HST for the nearest globular clusters. It might
also be possible to find black holes in globular clusters by their gravitational wave emission. The detection
through the X-ray emission arising from the IMBH is possible only after the tidal disruption of a star
or when a star captured through tidal heating is in a close enough orbit to the IMBH. 
Intermediate mass black holes also speed up the dissolution
of star clusters if the clusters are surrounded by a tidal field.


\end{document}